\documentstyle[12pt]{article}

\def\mfor{{\hspace{0.3cm} \rm for} \hspace{0.3cm}}
\def\mpunkt{{\hspace{0.3cm} \rm .}}

\begin{document}
\textheight23.cm
\textwidth15cm
\renewcommand{\baselinestretch}{1.2}
\title{Numerical Investigation of the Evolution of Sandpiles}
\author{Volkhard Buchholtz and Thorsten P\"oschel}
\date{\small \today}
\maketitle
\begin{center}
        preprint HLRZ 56/93 \\
        \vspace*{10ex}
	HLRZ, KFA J\"ulich, Postfach 1913 \\
	D--52425 J\"ulich, Germany  \\
 		\vspace*{1ex}
	and \\
 		\vspace*{1ex}
	Humboldt--Universit\"at zu Berlin \\
	FB Physik, Institut f\"ur Theoretische Physik \\
	Invalidenstra\ss e 42, D--10115 Berlin \\
\end{center}
\vspace*{2ex}
\begin{abstract}
\noindent
The evolution of a pile of granular material is investigated by
molecular dynamics using a new model including nonsphericity of the
particles instead of introducing static friction terms. The angle of
repose of the piles as well as the avalanche statistics gathered by
the simulation agree with experimental results. The angle of repose of
the pile is determined by the shape of the grains. Our results are
compared with simulations using spherical grains and static friction.
\end{abstract}
\vfill
\setcounter{topnumber}{5}
\renewcommand{\topfraction}{0.99}
\setcounter{bottomnumber}{5}
\setcounter{totalnumber}{5}
\renewcommand{\bottomfraction}{0.99}
\renewcommand{\textfraction}{0.01}
\newpage
\noindent
\renewcommand{\baselinestretch}{1.4}
\small \normalsize
\section{Introduction}
There are many interesting phenomena observed in the dynamics of
granular materials which have been studied by many authors and by
various methods (e.g.~\cite{jaeger_nagel_science}\cite{herrmann} and
references therein). One of the most interesting phenomena is the
evolution of a pile of granular material. When sand grains are dropped
on the top of a pile its slope does not depend on the number of grains
as long as the heap is not too small. The avalanches going down the
surface of the heap have no characteristic size, they are only limited
by the size of the entire pile. The size of the avalanches as well as
the time interval between successive avalanches fluctuate irregularly.
Their spectra decay possibly due to a power
law~\cite{rosendahl}--\cite{kadanoff}.
On large scales, however, there was measured a different
behaviour~\cite{jaeger}. Recently Rosendahl~et.~al.~\cite{rosendahl}
found that the predominant number of avalanches is
power--law--distributed but that these avalanches are supplemented by
periodic significantly larger avalanches. Dependent on the material
the slope of the surface of a heap can vary, hence one observes waves
creeping up the pile~\cite{frette}.
\par
There are different numerical methods to simulate the static and
dynamic behaviour of macroscopic amounts of granular media. Cundall
and Strack~\cite{cundall} introduced a model for the interaction of
grains which is widely used in molecular dynamics simulations. Other
simulations base on cellular automata~\cite{baxter}, the results led
to the idea of self organized criticality~\cite{bak}. Other numerical
investigations base on the Boltzmann lattice gas~\cite{flekkoy}, on
Monte Carlo simulations (e.g.~\cite{devillard}\cite{rosato}) or on
random walk models~\cite{caram}.
\par
In this paper we apply a new model for the simulation of the evolution
of sandpiles using molecular dynamics as introduced
in~\cite{poeschel_buchholtz}. Many of the numerical investigations
base on molecular dynamics
(e.g.~\cite{herrmann}\cite{cundall}\cite{haff}--\cite{poeschel}),
however, most of them use spherical grains. To simulate static
friction between the grains one introduces a force which is due to the
Coulomb law, i.e. the spheres slide on each other for the case that
the shear force between the grains $F_S$ is larger than $\mu \cdot
F_N$ where $F_N$ is the force in normal direction, $\mu$ is the
Coulomb friction coefficient. This friction force is derived from the
phenomenological idea about friction between macroscopic bodies. Its
microscopic origin is still hardly
understood~\cite{poeschel_herrmann}. In our molecular dynamics
simulations we use nonspherical particles instead of spheres. The
particles we use here are elastic but not stiff, they deform during
collisions and dissipate energy.
\par
Recently Gallas and Soko\l owski suggested a model for nonspherical
grains where two spheres were connected by a stiff
bar~\cite{gallas_s}. Their results agree qualitatively with ours.
\par
As we will demonstrate below the results of the simulations using
nonspherical grains agree better with experimental investigations than
traditional models.
\section{The Model}
In our simulation each of the grains $k$ consists of five spheres,
four of them with radii $r_i^{(k)}$ are located at the corners of a
square of size $L^{(k)}$. The fifth sphere is situated in the middle
of the square, its radius is chosen to touch the others
\begin{equation}
r_m^{(k)}=\frac{L^{(k)}}{\sqrt{2}}-r_i^{(k)} \mpunkt
\end{equation}
Each of the spheres is connected to all of its neighbouring spheres
which belong to the same grain by damped springs (fig.~\ref{geopart}).
\begin{figure}[ht]
\caption{\it Each of the particles $k$ consists of four spheres with
equal radii $r_i^{(k)}$ at the corners of a square of size $L^{(k)}$
and one sphere in the centre. Its radius is chosen to touch the
others. There are forces due to a damped spring acting between the
spheres the grain consists of.}
\label{geopart}
\vspace{2ex}
\end{figure}
\par
\noindent
\par
Between each two spheres $i$ and $j$ which might belong to the same
particle or to different particles acts the force $\vec{F}_{ij}$:
\begin{equation}
\vec{F}_{ij} =  \left\{ \begin{array}{cl}
	F_{ij}^N \cdot
	\frac{\vec{x}_i - \vec{x}_j}{|\vec{x}_i-\vec{x}_j|}\hspace{0.5cm}
		& \mbox{if $|\vec{x}_i-\vec{x}_j| < r_i + r_j$}
		\\[1ex]
		0
			& \mbox{otherwise}
		\end{array}
		\right.
\end{equation}
with the normal force
\begin{equation}
F_{ij}^N=Y \cdot (r_i+r_j-|\vec{x}_i-\vec{x}_j|)~-~\gamma \cdot
	 m_{ij}^{eff}\cdot |\dot{\vec{x}}_i - \dot{\vec{x}}_j|
\end{equation}
and the effective mass
\begin{equation}
m_{ij}^{eff} = \frac{m_i \cdot m_j}{m_i + m_j} \mpunkt
\end{equation}
$Y$ and $\gamma$ are the Young modulus characterizing the elastic
restoration of the spheres and the phenomenological damping
coefficient. The terms $\vec{x}_i$, $\dot{\vec{x}}_i$ and $m_i$ denote
the current position, velocity and mass of the $i$--th sphere. The
normal force $F_{ij}^N$ consists of an elastic restoration force which
corresponds to the microscopic assumption that the particles can
slightly deform each other and a term describing the energy
dissipation of the system due to collisions between particles
according to normal friction and plasticity.
\par
Moreover each pair of neighbored spheres $i$ and $j$ which both belong
to the same grain $k$  feel the force
\begin{equation}
\vec{F}_{ij}^{Sp} = \Big[ \alpha \cdot (C^{(k)}_{ij} - |\vec{x}_i-\vec{x}_j|) -
	\gamma_{Sp} \cdot \frac{m_i}{2} \cdot |\dot{\vec{x}}_i - \dot{\vec{x}}_j|
	\Big] \cdot
	\frac{\vec{x}_i - \vec{x}_j}{|\vec{x_i}-\vec{x}_j|}
\end{equation}
with
\begin{equation}
C^{(k)}_{ij} = \left\{ \begin{array}{ll}
		L^{(k)} & \mbox{if $i$, $j$ lie at the same edge}\\
		L^{(k)}/\sqrt{2} & \mbox{else} \mpunkt
		\end{array} \right.
\end{equation}
This force is due to in internal damping with spring constant $\alpha$
and damping coefficient $\gamma_{Sp}$ acting between particles of the
same grain.
\par
To compare our results using the nonspherical grains defined above we
describe now how to introduce static friction in the simulation using
spheres as it was done first by Cundall and Strack~\cite{cundall} and
modified by Haff and Werner~\cite{haff}. Here each pair of particles
$i$ and $j$ with radii $r_i$ and $r_j$ feel the force
\begin{equation}
	\vec{F}_{ij} =  \left\{ \begin{array}{cl}
		F_{ij}^N \cdot \frac{\vec{x}_i - \vec{x}_j}{|\vec{x_i}-\vec{x}_j|}
		+ F_{ij}^S \cdot  \left({0 \atop 1} ~{-1 \atop 0} \right) \cdot
                  \frac{\vec{x}_i - \vec{x}_j}{|\vec{x_i}-\vec{x}_j|}
			& \mbox{if $|\vec{x_i}-\vec{x}_j| < r_i + r_j$}
		\\[1ex]
		0
			& \mbox{otherwise}
		\end{array}
		\right.
\end{equation}
with the force in normal direction
\begin{equation}
	F_{ij}^N  = Y \cdot (r_i + r_j - |\vec{x}_i - \vec{x}_j|)~+
	~\gamma \cdot m_{ij}^{eff}\cdot |\dot{\vec{x}}_i - \dot{\vec{x}}_j|
\end{equation}
and the shear force
\begin{equation}
	F_{ij}^S = \min \{- \gamma_S \cdot m_{ij}^{eff} \cdot |\vec{v}_{ij}^{rel}|~,~
\mu \cdot
	|F_{ij}^N| \} \mpunkt
\label{eq_coulomb}
\end{equation}
The relative velocity of the particle surfaces results from the
relative velocity of the particles and their angular velocities
$\dot{\Omega}_i$ and
$\dot{\Omega}_j$
\begin{equation}
	\vec{v}_{ij}^{rel} = (\dot{\vec{x}}_i - \dot{\vec{x}}_j) + r_i \cdot
\dot{\Omega}_i + r_j \cdot \dot{\Omega}_j \mpunkt
\end{equation}
\vspace{2ex}
\par
\noindent
The parameters $\gamma$ and $\gamma_S$ stand for the normal and shear
friction coefficients, $\mu$ is the Coulomb friction parameter.
Eq.~(\ref{eq_coulomb}) takes into account that two grains slide upon
each other when he shear force between them exceeds a certain value,
otherwise they roll (Coulomb relation).
\par
The sizes of the nonspherical particles $L^{(k)}$ as well as the radii
$r_i$ in the case of spherical grains were Gauss distributed with mean
value $\overline{L^{(k)}}$.
\section{Results}
In our simulation we investigate the evolution of a sand pile by
consecutively dropping nonspherical particles on the top of the heap.
The surface of the platform upon which the heap is built up consists
of spheres with radii $r_i^{(p)}$ with mean value
$\overline{r_i^{(p)}}=\overline{L^{(k)}}$ to simulate a rough surface.
After each dropping a particle on the heap we waited until the
velocities of all particles of the entire heap faded away, i.e. until
they were smaller than a very small constant.
\par
For the parameters we used in the simulations: $Y=10^4 kg/s^2$,
$\gamma=1.5\cdot 10^4~s^{-1}$, $\alpha=10^4~kg/s^2$,
$\gamma_{Sp}=3\cdot 10^4 ~s^{-1}$, $L^{(k)}/r_i^{(k)}=4$,
$\overline{L^{(k)}}=3~mm$. The results were compared with simulations
using spherical grains, here we assumed the same values and
$\gamma_S=3\cdot 10^{4}~s^{-1}$, $\mu =0.5$ and $\overline{r_i}=3~mm$.
\par
For the calculation of the evolution of the particle positions we
applied a Gear predictor corrector algorithm~\cite{predictor} of
sixth order. For the case of the simulation using spherical grains we
have further to calculate the rotation of each sphere. Since the
acceleration of the particle surface due to particle rotation is much
smaller than due to particle movement we applied therefore a Gear
algorithm of  fourth order only.
\subsection{Angle of Repose}
When building up a heap experimentally one finds  that the slope
$\Phi$ (angle of repose) does not depend on the particle number. It
depends on the raw material and lies typically between $20^o$ and
$30^o$. Bretz et.~al. found $\Phi \approx 25^o$~\cite{bretz}.
\par
Figure~\ref{haufen} shows snapshots of simulated piles of different
sizes. The size of the piles have been scaled so that the widths of
all piles appear equal in the figures. The first two piles consist of
$N=300$ (a) and $N=1100$ (b) nonspherical particles. The slope is
approximately the same for both figures, i.e. it does not depend on
the number of particles. Figs. (c) and (d) show piles of $N=400$ and
$N=1800$ spherical particles. In the case of spherical grains the heap
dissolves under gravity, the slope depends on the number of particles.
\begin{figure}[ht]
\caption{\it Snapshots of simulated  piles. Heaps (a) and (b) consist
of $N=300$ and $N=1100$ nonspherical grains, heaps (c) and (d) of
$N=400$ and $N=1800$ spherical particles. In the latter case the slope
depends on the number of particles.}
\label{haufen}
\vspace{2ex}
\end{figure}
\par
In the simulations using nonspherical grains we measured $\Phi \approx
21^o$, this value is in agreement with experimentally found
data~\cite{bretz}. Fig.~\ref{winkel_anzahl} shows the evolution of the
slope $\Phi$ of the pile during one run over the particle number for
spherical and nonspherical grains. For nonspherical grains the slope
$\Phi$ fluctuates due to avalanches of different size. Except for very
small heaps the average slope does not depend on the particle number
in accordance with the experiment. For the case of spherical particles
the slope decays with rising particle number.
\begin{figure}[ht]
\caption{\it The evolution of the slope of the pile during one run
over the particle number for both spherical and nonspherical grains.
For nonspherical grains the mean angle does not depend on the particle
number while it decreases for spherical particles.}
\label{winkel_anzahl}
\vspace{2ex}
\end{figure}
\par
Since the surface of the pile is not a smooth line we need a special
procedure to calculate the slope $\Phi$. Fig.~\ref{winkelmessen} shows
a pile with an arbitrarily shaped heap. The shape of the pile is
described by the function $h(x)$ with $x\in [0,x_{max}]$, $x_{cm}$ and
$y_{cm}$ denote the coordinates of the centre of mass of the heap. The
height $H$ of an ideal triangle which has the same baselength and
volume like the heap would be
\begin{equation}
H=\frac{2}{x_{max}}\cdot \int\limits_0^{x_{max}}h(x)\,dx \mpunkt
\end{equation}
Because the centre of mass of a triangle is the intersection point of
all three median lines we calculate the baselength $B$ of the triangle
using the intercept theorem
\begin{equation}
\frac{2 \cdot H}{B} = \frac{H-y_{cm}}{x_{cm}} \mpunkt
\end{equation}
Hence we find for the slope
\begin{equation}
\Phi=\arctan \left(\frac{H-y_{cm}}{2\cdot x_{cm}} \right) \mpunkt
\label{phi}
\end{equation}
Eq.~(\ref{phi}) gives a good approximation for the slope provided that
the shape of the heap is close to a triangle. The snapshots in
fig.~\ref{haufen} show that this precondition is given in our case.
\par
Notice that there are two possibilities of defining the angle~\cite{lee},
the angle of repose and the angle of marginal stability. For
simplicity we used in this work the time average of the angle observed
during the simulation.
\begin{figure}[ht]
\caption{\it Schematic plot of the procedure used to determine the
slope of a pile which surface is not a smooth plane.}
\label{winkelmessen}
\end{figure}
\subsection{Avalanches and Mass Fluctuations}
Theoretical as well as experimental
investigations~\cite{rosendahl}--\cite{kadanoff}\cite{bak}
led to the hypothesis that the size of the avalanches, i.e. the mass
fluctuations of the pile as well as the distribution of the time
intervals between successive avalanches of sandpiles can be described
by the self organized criticality--model. In order to investigate the
size of the avalanches and the distribution of the time between each
two successive avalanches we modified the setup of the simulation and
monitor the fluctuation of the mass $m_h$ of a heap of definite size.
Instead of an infinite platform we use a platform of a finite length
$P$ above which the heap is built up. In fig.~\ref{masse_zeit} is
drawn a part of the time series of the mass $m_h$ for fixed $P=820~mm
\approx 273~\overline{L^{(k)}}$. The mass fluctuates irregularly due
to avalanches of different size going down the surface of the heap.
Since after dropping a grain on the heap we wait until the motion of
all grains fades away the natural unit for measuring the time is the
number of dropping events, but not the number of integration steps
used in the simulation. Since we want to make statistics for all
avalanches but not only for those which cause a change of the mass
$m_h$ of the heap we monitor the change of the slope of the heap, i.e.
the change of the centre of mass, during an avalanche instead of the
change of the mass. Otherwise we would cut off the lower part of the
spectrum, most of the small avalanches do not reach the bottom of the
pile. Fig.~\ref{fluct} shows the time series of the avalanches.
Fig.~\ref{fluct}a displays the changes of the slope, fig.~\ref{fluct}b
the changes of the mass. Both figures resemble each other but in the
latter case many of the small avalanches have been cut off. This is
due to small avalanches which do not reach the bottom and therefore do
not change the mass of the heap. The log--log--plot of the spectrum
(fig.~\ref{log_log_size}) shows that the size distribution of the
avalanches might follow a power law. For the exponent $h(N_A) \sim
(N_A)^m$ we measured $m\approx -1.4$. In the experiments was found
$m\approx -2.5$~\cite{held} and $m\approx -2.134$~\cite{bretz}
respectively.
\begin{figure}[ht]
\caption{\it Total mass of a pile of nonspherical grains on a platform
of finite length $P = 820~mm \approx 273~\overline{L^{(k)}}$. The mass
fluctuates irregularly due to avalanches of different size. The time
is measured in dropping events.}
\label{masse_zeit}
\end{figure}
\begin{figure}[ht]
\caption{\it Time series of the changes of the slope (a) and of the
total mass (b) of the heap due to avalanches. Many of the small
avalanches do not cause a change of the mass of the heap, since they
stop before reaching the bottom.}
\label{fluct}
\end{figure}
\begin{figure}[ht]
\caption{\it Size distribution of the avalanches. The line displays
the function $h(N_A)\sim (N_A)^{-1.4}$.}
\label{log_log_size}
\end{figure}
The spectrum of the time intervals between each two successive
avalanches was experimentally investigated too.
Fig.~\ref{spectrum_time} shows the spectrum of the distribution of the
time intervals between two successive avalanches which we found in the
simulation. We are not convinced that the decay of the spectrum in
our simulation follows a power law too.
\begin{figure}[ht]
\caption{\it Spectrum of the time intervals between successive
avalanches.}
\label{spectrum_time}
\end{figure}
\subsection{Influence of the Particle Sphericity}
Dependent on the ratio $L^{(k)}/r_i^{(k)}$ the grain $k$ is more
similar to a square or to a sphere. Hence we may investigate the
influence of the particle shape on the angle of repose of a pile. We
define the shape $S$ of a grain
\begin{equation}
S=1-\frac{R_{min}^{cc}}{R_{max}^{cc}}
\end{equation}
where $R_{min}^{cc}$ and $R_{max}^{cc}$ are the extremal values of the
distance between the convex cover of the nonspherical grain and its
central point (fig.~\ref{exc_fig}). In terms of $L^{(k)}/r_i^{(k)}$ we
write
\begin{eqnarray}
R_{min}^{cc} & = & \max \left( \frac{1}{\sqrt{2}} \frac{L^{(k)}}{r_i^{(k)}} -
	1,\,\, \frac{1}{2} \frac{L^{(k)}}{r_i^{(k)}} + 1 \right) \\
R_{max}^{cc} & = & \frac{1}{\sqrt{2}} \frac{L^{(k)}}{r_i^{(k)}} + 1
\end{eqnarray}
\begin{figure}[ht]
\caption{\it The shape $S$ is determined by the extreme sizes of the
convex cover of the grains $R_{min}^{cc}$ and $R_{max}^{cc}$.}
\label{exc_fig}
\end{figure}
Fig.~\ref{s} displays the shape value $S$ as a function of
$L^{(k)}/r_i^{(k)}$. The function $S$ reaches its maximum $S_m$ for a
particle which convex cover is most similar to a square
($(L^{(k)}/r_i^{(k)})_{S_m}=9.66$). For each $S<S_m$ there are two
different particle shapes with the same shape value. Since
$L^{(k)}/r_i^{(k)}\ge 2$ there are no particles with $S<0.172$ for
lower $L^{(k)}/r_i^{(k)}$ ratio. For the limit $L^{(k)}/r_i^{(k)}
\rightarrow \infty$ the grains are spheres and $S$ becomes zero.
\begin{figure}[ht]
\caption{\it The shape value $S$ as a function of $L^{(k)}/r_i^{(k)}$.
Each value $S$ with $S<S_{m}$ corresponds to two different particle
shapes with the same shape value, one with lower and one with larger
$L^{(k)}/r_i^{(k)}$ ratio.}
\label{s}
\end{figure}
\par
To investigate the influence of the shape $S$ of the grains on the
result of the simulations we have to scale the density of the material
the grains consist of to ensure that the total mass $m^{(k)}$ of each
grain remains constant independent on its shape ($\rho_0 \rightarrow
\rho$ when $S_0 \rightarrow S$, $L^{(k)}=const.$, $m^{(k)}=const.$)
\begin{equation}
\rho(S) = \left\{ \begin{array}{ll}
\rho_0 \cdot \frac{(2-S)^2}{(2-S_0)^2} \frac{4S_0^2-4S_0+2}{4S^2-4S+2} &
\mfor \frac{L^{(k)}}{r_i^{(k)}} \ge \left(
\frac{L^{(k)}}{r_i^{(k)}}\right)_{S_m} \vspace{0.2cm}\\
\rho_0 \cdot \frac{4S^2}{4S_0^2} \frac{16S_0^2+\left( 12\sqrt{2} - 24
\right)S_0 + 15
-10\sqrt{2}}
	{16S^2+\left( 12\sqrt{2} - 24 \right)S + 15 -10\sqrt{2}} &
\mfor \frac{L^{(k)}}{r_i^{(k)}} \le \left(
\frac{L^{(k)}}{r_i^{(k)}}\right)_{S_m} \mpunkt
\end{array} \right.
\end{equation}
The influence of the shape of the grains to the slope of the heap is
drawn in fig.~\ref{winkel_excentr}. As mentioned above each $S$
corresponds to two arguments $L^{(k)}/r_i^{(k)}$. Therefore we denote
the slope for grains with $L^{(k)}/r_i^{(k)} \le
(L^{(k)}/r_i^{(k)})_{S_m}$ by $\odot$ and by $+$ for
$L^{(k)}/r_i^{(k)} \ge (L^{(k)}/r_i^{(k)})_{S_m}$. As expected the
slope $\Phi$ reaches its maximum for particles with maximum shape
value $S=S_m$ since they are most similar to squares. The dashed line
marks the inclination $\Phi_{sp}$ we found for a heap of spheres,
which corresponds to $S \rightarrow 0$. The slope values we measured
simulating nonspherical grains lie between the slope of a pile of
spheres and a pile of grains with $S=S_m$.
\begin{figure}[ht]
\caption{\it Slope $\Phi$ of a heap over the shape value $S$ for
grains with $L^{(k)}/r_i^{(k)} \le (L^{(k)}/r_i^{(k)})_{S_m}$
$(\odot)$ and $L^{(k)}/r_i^{(k)} \ge (L^{(k)}/r_i^{(k)})_{S_m}$ $(+)$.
The dotted line leads the eye to the function $\Phi = 130 \cdot S +
const$. The dashed line displays the inclination observed in
simulations with spherical particles.}
\label{winkel_excentr}
\end{figure}
\par
The calculation for the data shown in fig.~\ref{winkel_excentr} are
extremely time consuming. For this reason we are not able to present
more data.
\section{Conclusion}
Using molecular dynamics simulations we investigated the evolution of
a sand pile and the statistics of the avalanches going down. Simulated
piles using nonspherical grains remain stable under gravity,
independent on their size the slope is approximately the same. In  the
simulation we found for the average slope $\overline{\Phi} \approx
13^o \dots 23^o$ dependent on the shape of the grains. In experiments
was found $\overline{\Phi} \approx 25^o$~\cite{bretz}. For the case of
simulations using spheres and friction forces due to the Coulomb law
the slop of the heap decreases with increasing particle number.
\par
During the simulation the mass of the heap which is built up upon a
finite platform varies irregularly. For the statistics of the
avalanches we used the fluctuations of the slope of the heap, using
the fluctuations of the mass leads to wrong results since most of the
small avalanches do not reach the bottom of the heap and therefore do
not change its mass. In our simulations the avalanches are
power--law--distributed, for the exponent we found $h(N_A)\sim
(N_A)^{-1.4}$.
\par
Our model allows for changing the shape of the grains. We investigated
the influence of their shape to the slope of the heap of constant
size. We found that the slope reaches its maximum for grains which
convex cover is most similar to a square. The minimum slope appears
for the case that the particles are spheres. The slope of piles built
of other nonspherical particles lies in between the minimum and the
maximum.
\section{Acknowledgment}
We thank J.~A.~C.~Gallas and H.~J.~Herrmann for helpful discussions.

\newpage
\listoffigures
\end{document}